\documentclass[11pt,twocolumn,aps,prb,reprint]{revtex4-1}

\usepackage{braket}   
\usepackage{graphicx,amsmath,amssymb,braket}

\begin{document}

\title{Hydrodynamic-to-ballistic crossover in Dirac fluid}
\author{D. Svintsov}
\affiliation{Laboratory of 2d Materials' Optoelectronics, Moscow Institute of Physics and Technology, Dolgoprudny 141700, Russia}
\affiliation{Institute of Physics and Technology, Russian Academy of Science, Moscow 117218, Russia}

\begin{abstract}
We develop an analytically solvable classical kinetic model of spatially dispersive transport in Dirac materials accounting for strong electron-electron (e-e) and electron-hole (e-h) collisions. We use this model to track the evolution of graphene conductivity and properties of its collective excitations across the hydrodynamic-to-ballistic crossover. We find the relaxation rate of electric current by e-e collisions that is possible due to the lack of Galilean invariance, and introduce a universal numerical measure of this non-invariance. We find the two branches of collective excitations in the Dirac fluid: plasmons and electron-hole sound. The sound waves persist at frequencies exceeding the e-e collision frequency, have a small viscous damping at the neutrality point, but acquire large damping due to e-h friction even at slight doping. On the contrary, plasmons acquire strong frictional damping at the neutrality point and become well-defined in doped samples.
\end{abstract}

\maketitle
In ultra-clean semiconductor heterostructures~\cite{High-mobility,Banszerus}, the frequency of electron-electron collisions $\tau_{ee}^{-1}$ can exceed than the frequency of electron-phonon $\tau^{-1}_{\rm e-ph}$ and electron-impurity $\tau^{-1}_{\rm e-i}$ collisions. This results in hydrodynamic (HD) regime of electron transport governed by Navier-Stokes equation~\cite{gantmakher2012carrier,Gurzhi-HD,deJong-HD}, which contrasts to the diffusive transport governed by Ohm's law. The hydrodynamic regime can potentially span up to terahertz frequencies~\cite{Knap-resonant,bandurin2017dual} of driving field $\omega$, until it surrenders to the balilistic motion of individual quasiparticles at $\omega \tau_{ee} \gg 1$. Several bright confirmations of HD transport appeared recently for Dirac materials with linear carrier dispersion, including the observation of whirlpools~\cite{Bandurin-HDgraphene} and Poiselle flow in graphene~\cite{Bandurin-superballistic} and thermoelectric signatures of axial-gravitational anomaly in Weyl semimetals~\cite{Gooth2017}.

In most experimentally relevant situations, however, the frequencies $\tau_{ee}^{-1}$, $\tau^{-1}_{\rm e-ph}$ and $\omega$ are of the same order of magnitude, which results in the transport regime intermediate between hydrodynamic~\cite{Muller_nearly_perfect,Muller_magnetotransport,Mirlin-HD,Foster-Aleiner,Our-hydrodynamic,Lucas_HDWeyl}, diffusive~\cite{Vasko-Ryzhii,AdamSelf-Consistent}, and ballistic~\cite{Ballistic-graphene}. For collective excitations in Dirac materials, the transport is also highly {\it nonlocal}~\cite{Lundeberg_Nonlocal_Plasmons} implying that their characteristic wavelength $\lambda$ is comparable to $v_0/\omega$, where $v_0$ is the Fermi velocity.

The studies of intermediate transport regimes are required not only to describe the experimental reality, but also to sort out the conflicting predictions of hydrodynamic and ballistic models. The HD theory for Dirac materials predicts that e-e collisions affect conductivity only indirectly by modifying the distribution function and relaxation rates~\cite{MacDonaldHD}. The model of nearly free particles tells that e-e collisions can relax the net current~\cite{Kashuba_ConductivityDefectless,Fritz_PRB,Sun_EE_current_relaxation,CC-scattering-NDC} due to the absence of direct proportionality between momentum and velocity. The free-particle models tell that the phase velocity of collective excitations -- plasmons -- always lies above the Fermi velocity~\cite{Ryzhii-plasmons,Das_Sarma_Plasmons,DasSarma_PlasmWeyl}, which protects them both from Landau damping~\cite{Lundeberg_Nonlocal_Plasmons} and current-driven instabilities~\cite{Stauber-gain}. The HD model sets a lower bound for plasmon velocity to $v_0/\sqrt{d}$, where $d$ is the dimension of space~\cite{Lucas_SoundWaves,Polini_FET}. Remarkably, $v_0/\sqrt{d}$ is the velocity a novel quasi-neutral hydrodynamic mode~\cite{Our-hydrodynamic,Levitov_EnergyWaves,Lucas_SoundWaves}, which does not appear in the quantum analysis~\cite{Das_Sarma_Plasmons} even after inclusion of electron interactions~\cite{Abedinpour_NonGalilean,Plasmons-interactions}.

In this paper, we develop an exactly solvable kinetic model describing {\it nonlocal} ballistic and hydrodynamic regimes in the materials with linear carrier dispersion $\epsilon_{\bf p} = \pm v_0 p$, and resolve the abovementioned problems. We find a closed-form expression for dispersion law of collective excitations (plasmons and electron-hole sound) in graphene valid at frequencies and wave vectors unaccessible by preceding models. Its solution demonstrates a gradual increase in plasmon velocity from HD to ballistic regime. In the intermediate regime, $\omega \tau_{ee}\sim 1$, we find the maximum damping of plasmons solely due to electron-electron scattering that corresponds to the quality factor $Q\sim 10...30$. We find that electron-hole sound surprisingly persists both in HD and ballistic limits; in the latter case its velocity tends to $v_0$ from the lower side. We also identify a new origin of damping for collective excitations due to energy momentum transfer between electrons and holes. The latter is relevant for plasmons at charge neutrality and sound in doped samples.

The theories of intermediate transport regimes rely either on solution of full kinetic equation in a restricted basis~\cite{Muller_magnetotransport,Sun_LinearNonlinear}, or on the replacement of true e-e collision integrals with simple model forms~\cite{Abrikosov-Khalatnikov,BGK-collisions,BGK-collisionsII,Mermin-Lindhard_dielectric_Function,lucas2017kinetic}. The latter approach was previously applied to the conductivity of graphene nano-constrictions~\cite{Levitov_higher-than-ballistic} and circular obstacles~\cite{Stokes-paradox}, and to the problem of viscous electron-hole sound damping~\cite{Levitov_higher-than-ballistic}. In this paper, we  extend the applicability of model collision integral approach to strong nonlocality,  finite temperatures, and electron-hole scattering.

The transport coefficients of electron-hole system in a weak electric field ${\bf E} \propto e^{i({\bf qr}-\omega t)}$ can be found from solution of coupled kinetic equations for distribution functions $ f_{e}$ and $f_h$:
\begin{equation}
\label{Kinetic}
-i(\omega - {\bf q v}_{\bf p} ) f_{e}  - e {\bf E} {\bf v}_{\bf p} \frac{\partial f_{0e}}{\partial \epsilon_{\bf p}} = \mathcal{C}_{ee}[f_{e}] + \mathcal{C}_{eh}[ f_{e},f_{h}],
\end{equation}
a similar equation with opposite charge and $e$ and $h$ indices interchanged holds for holes. Here ${\bf v}_{\bf p} = \partial \epsilon_{\bf p}/\partial{\bf p}$ is the quasiparticle velocity, $f_{0e}$ is the distribution function in the absence of field. Below we supplement steady-state quantities with subscript 0.

E-e collisions relax the perturbations of the distribution function to the local-equilibrium
\begin{equation}
\label{HD_function}
f^{\rm hd}_{e} = -\frac{\partial f_{0e}}{\partial \epsilon_{\bf p}}\left(\mu_e + {\bf p}{\bf u}_e +\xi_{\bf p} \frac{T_e}{T_0}   \right),
\end{equation}
where $\mu_e$, ${\bf u}_e$, and $T_e$ are the small perturbations of Fermi energy, drift velocity and temperature, and $\xi_{\bf p}$ is the energy reckoned from the Fermi energy. Each of the terms in Eq.~(\ref{HD_function}) represents a zero mode of the true e-e collision operator. The model collision operator should have the same zero modes and the simplest possible form:
\begin{equation}
\label{Cee}
\tilde{\mathcal{C}}_{ee} = - \frac{ f_e - f^{\rm hd}_{e}}{\tau_{ee}}.
\end{equation}

Collisions of electrons with holes also lead to local equilibrium, but its parameters depend on both electron and hole Fermi energy, drift velocity, and temperature~\cite{BGK-collisionsII}. These dependences are modeled to ensure the conservation of {\it total} electron-hole momentum and energy and {\it separate} electron and hole numbers. In addition, there should be no momentum (energy) transfer between electrons and holes if their drift velocities (temperatures) are equal. Within the linear response, there is an only one possible form of model collision integral satisfying these properties (Supporting information, sec. I):
\begin{multline}
\label{Electron-hole}
{{\tilde{\mathcal C}}_{eh}}
=-\frac{f_e -f^{hd}_{e} }{\tau_{eh}}+\\
\frac{\partial f_{0e}}{\partial \varepsilon }
\left[ \frac{ {\bf p}\Delta {\bf u}_{eh}}{\tau^{\rm tr}_{\rm eh}}+\frac{\Delta T_{eh}}{{\tau^{\varepsilon}_{\rm eh}}}\left( \frac{\xi_p}{T_0} - \frac{\partial n_e/\partial T }{\partial n_e/\partial \mu} \right) \right],
\end{multline}
where $\Delta {\bf u}_{eh} = {\bf u}_{e} - {\bf u}_{h}$, $\Delta T_{eh} = T_{e} - T_{h}$. Equation~(\ref{Electron-hole}) explains that e-h collisions tend the electron distribution function to a local equilibrium, while electron and hole parameters are pushed toward each other. 

The scattering times $\tau_{ee}$ and $\tau_{eh}$ can be estimated from imaginary part of electron self-energy, while transport and energy times ${\tau^{tr}_{\rm eh}}$ and ${\tau^{\varepsilon}_{\rm eh}}$ are found from microscopic calculations of momentum and energy exchange between electrons and holes (Supporting information, sec. II). Proceeding this way, we find $\tau^{-1}_{ee} \sim 5 \times 10^{12}$ s$^{-1}$ at room temperature and $\mu = 0...4 T$, $\tau^{-1}_{eh} \approx 10^{13}$ s$^{-1}$ at the neutrality point and falls exponentially away from it (see Fig.~\ref{Figure_FrictionalDamping}). This justifies the hydrodynamic approach up to terahertz frequencies.

The closure of kinetic model is achieved by requiring the particle number, momentum, and energy conservation upon e-e and h-h collisions. This yields six hydrodynamic equations for carrier densities $n_\alpha$, momenta $\rho_{0\alpha}u_\alpha$, and energy densities $\varepsilon_\alpha$ ($\alpha = \{e,h\}$, $\rho_{0\alpha} = (d+1)\varepsilon_{0\alpha}/dv_0^2$ is the effective mass density). Their formulation at arbitrary wave vector $q$ is possible due to the independence of carrier velocity on momentum modulus. In this case, the energy dependence of modes excited in non-uniform field is the same as of hydrodynamic modes (\ref{HD_function}). As a result, the hydrodynamic equations acquire  a simple form:
\begin{gather}
\label{Gen-continuity}
[-i\omega +\tau_{e}^{-1} (1-\mathcal I_0) ] n_e - \frac{\mathcal I_1}{v_0\tau_e}{j^{(n)}_{e}}=0, \\ 
\label{Gen-Euler}
[ -i\omega +\tau_{e}^{-1} (1-\mathcal I_2) ]\rho_{0e} u_e -\frac{\mathcal I_1}{v_0\tau_e} {\Pi_e}=F_e, \\ 
\label{Gen-energy}
[ -i\omega +\tau_{e}^{-1}(1-\mathcal I_0) ]\varepsilon_e - \frac{\mathcal I_1}{v_0\tau_e}{j^{(\varepsilon)}_e}=Q_e, 
\end{gather}
where $j^{(n)}_{e}$ and $j^{(\varepsilon)}_{e}$ are the particle and energy currents, $\Pi_e$ is the pressure, and $\tau_e^{-1} = \tau_{ee}^{-1}+\tau_{eh}^{-1}$ is the net electron collision frequency. All information about spatial dispersion is encoded in dimensionless functions $\mathcal I_n$ depending on a single argument $w= (\omega + i \tau_e^{-1})/qv_0$:
\begin{equation}
{\mathcal I}_n(w) = \frac{1}{c_n} \left\langle{\frac{\cos^n\theta }{1 - w^{-1}\cos \theta}}\right\rangle,
\end{equation}
where $\langle...\rangle$ means the angular average, and $c_n = \langle\cos^{n+[1-(-1)^n]/2}\theta\rangle$. The functions $\mathcal I_n$ produce singularities in nonlocal response at $\omega\rightarrow qv_0$ in the absence of collisions (Supporting information, sec. III). For strong collisions, on the contrary, $\mathcal I_0 \approx \mathcal I_2 \approx 1$, $\mathcal I_1/v_0\tau_e \approx -i q$, and we restore an ordinary weakly non-local hydrodynamics~\cite{Muller_magnetotransport,Our-hydrodynamic}.

Currents and pressure in (\ref{Gen-continuity}-\ref{Gen-energy}) are given by
\begin{gather}
\label{NormalCurrent}
j^{(n)}_{e} = n_{0e}\left[  u_e - \frac{\tau_e}{\tau^{\rm tr}_{\rm eh}} \Delta u_{eh} - \frac{e\tau_e}{M_{e,k}} E \right],\\
\label{EnergyCurrent}
j^{(\varepsilon)}_{e} = \rho_{0e}\left[ u_e-\frac{\tau_e}{\tau^{\rm tr}_{\rm eh}}\Delta u_{eh} - \frac{e \tau_e}{M_{e,hd}} E \right],\\
\label{Pressure}
\Pi_e=\frac{1}{d}\left[{\varepsilon_e}-\frac{\partial \varepsilon_e}{\partial T}\frac{\tau_e}{\tau^{\rm th}_{\rm eh}}\Delta T_{eh}\right],
\end{gather}
they are renormalized by electron-hole scattering and contain dissipative corrections in external field similar to intrinsic charge conductivity in Ref.~\onlinecite{Muller_magnetotransport}. Above, it was convenient to introduce 'kinetic' and 'hydrodynamic' masses $M_{e,k}= d n_{0e} / (v_0^2 \partial n_{e}/\partial \mu)$ and $M_{e,hd} = \rho_{0e}/n_{0e}$. Electron-hole renormalization of currents can be viewed as mass enhancement due to scattering. Finally, $F_e$ and $Q_e$ are the force density and power transfer density
\begin{gather}
\label{Force}
F_e = -\mathcal I_2 \left[e n_{0e} E + \rho_{0e} \frac{\Delta u_{eh}}{\tau^{\rm tr}_{\rm eh}}\right],\\
\label{Heat}
Q_e = {\mathcal I_0}\frac{\partial \varepsilon_e}{\partial T}\frac{\Delta T_{eh}}{\tau^{\rm th}_{\rm eh}},
\end{gather}
they include the effect of external field and energy-momentum transfer between electrons and holes.

{\it Electron-electron scattering and conductivity.}
In the absence of spatial dispersion, the electron distribution function solving (\ref{Kinetic}) is a combination of hydrodynamic modes and the velocity mode $f_{e} \propto {\bf v_p} {\bf E} \partial f_{0e}/\partial\epsilon_{\bf p}$. In this case, predictions of our model should coincide with solutions of kinetic equation based on mode expansion~\cite{Muller_magnetotransport,Sun_LinearNonlinear}, and with extended hydrodynamic model including an imbalance mode~\cite{Mirlin-HD}. 

Particularly, the dc conductivity of graphene at the neutrality point $\sigma_{\mu=0}$ is found to be:
\begin{equation}
\sigma_{\mu=0} = e^2 n \left[\frac{2 \tau_{e}}{ M_k} + \frac{\tau^{tr}_{eh}}{M_{hd}}\left( 1 - \frac{2\tau_e}{\tau^{tr}_{eh}}\right) \right].
\end{equation}
The first term here is due to velocity modes which are damped both by e-e and e-h collisions~\cite{Fritz_PRB,Kashuba_ConductivityDefectless}. The second term comes from electron-hole hydrodynamic modes damped only by mutual scattering~\cite{Our-hydrodynamic}~\footnote{The first term of this result was obtained in~\cite{Fritz_PRB}. The authors argued that velocity modes have non-divergent eigenvalues, while other modes (particularly, a mode with opposite electron and hole velocities) have eigenvalues $\propto\ln \alpha_c$ due to collinear scattering anomaly. We note that latter eigenvalues are not indeed large as the coupling constant is order of unity, while the collinear scattering singularity is softened by screening.}.  Both terms yield conductivity scaling $\sigma_{\mu=0} \propto \alpha_c^{-2} e^2/\hbar$, where $\alpha_c$ is the coupling constant.

A similar ''two-mode'' result is obtained for high-frequency local conductivity in the  limit $\mu_0 \gg T$~\cite{Muller_CollectiveCyclotron,Sun_LinearNonlinear}
\begin{equation}
\label{EE-resistivity}
\sigma_{\mu\gg T} = i n_e e^2\left[ \frac{M^{-1}_{e,hd}}{ \omega} + \frac{M^{-1}_{e,k} - M^{-1}_{e,hd}}{ \omega + i \tau_{ee}^{-1}}  \right],
\end{equation}
where the first term is due to dissipationless hydrodynamic mode, and the second one - due to the velocity modes. At low frequencies, only the  hydrodynamic mode is excited, and resistivity of clean sample tends to zero. At high frequencies, the electromagnetic dissipation (${\rm Re}\sigma \neq 0$) appears only due to the lack of Galilean invariance, i.e. a subtle difference between ''kinetic'' and ''hydrodynamic'' masses. 

The emergence of small factor $1-M_{e,k}/M_{e,hd}$ in dissipative coefficients can be foreseen with the following arguments valid in arbitrary dimension. An electron fluid subjected to an electric field pulse of duration $\tau\ll \tau_{ee}$ acquires the momentum ${\bf p}_0$. The distribution of electrons after the pulse represents a shifted Fermi sphere $f_0(\epsilon_{{\bf p} - {\bf p}_0})$. The initial current density associated with this distribution is ${\bf j}_i = n_{0e} M^{-1}_{e,k} {\bf p}_0  $, the momentum density is ${\bf P} = n_{0e} {\bf p}_0$. E-e collisions establish a hydrodynamic distribution $f_0(\epsilon_{\bf p} - {\bf pu}_0)$ with the same total momentum ${\bf P} = n_{0e} M_{e,hd} {\bf u}_0$ but smaller current ${\bf j}_f ={\bf j}_i  M_{e,k}/M_{e,hd}$. The hydrodynamic mass is always less than kinetic mass, thus e-e collisions relax the current, but do not do it completely. The time dependence of current is ${\bf j}(t) = {\bf j}_f +({\bf j}_i - {\bf j}_f)e^{-t/\tau_{ee}} $, which allows one to introduce an effective {\it current relaxation} frequency
\begin{equation}
\label{Effective-relaxation}
\tau_{ee}^{*-1} = \tau^{-1}_{ee}\left[1 - M_{e,k}/M_{e,hd}\right].
\end{equation}

The relative difference of hydrodynamic and kinetic masses is thus the measure of Galilean non-invariance in Dirac fluid. In the degenerate limit, the invariance is restored, and the prefactor $1-M_{e,k}/M_{e,hd}$ goes to zero as $(T/\mu)^2$. The restoration of invariance~\cite{Principi_spin_charge_conductivity,Mirlin-HD} occurs because the electron velocity near the Fermi surface ${\bf v} \approx v_F {\bf p}/p$ is independent on dispersion law. 
As the quasiparticle lifetime is order of  $T^2/\mu_0$ by itself, the real part of ee conductivity in the high-frequency degenerate case scales as $T^4/\hbar^2 \omega^2 \mu_0^2$.

{\it Collective excitations at the ballistic-to-hydrodynamic crossover}. 
 The full power of derived equations (\ref{Gen-continuity}-\ref{Gen-energy}) is revealed for linear response in non-uniform fields, when the form of the distribution function is not known a priori and mode expansion/variational methods fail. The non-locality is crucial for description of collective excitations -- plasmons~\cite{Lundeberg_Nonlocal_Plasmons,Principi_plasmon_loss_hBN,Ryzhii-plasmons}.

 To study the collective modes in gated graphene, we supplement Eqs.~(\ref{Gen-continuity} -- \ref{Gen-energy}) with Gauss theorem $e \varphi = V_0(q)( n_h -  n_e)$, where $V_0(q)  = 2\pi e^2(1 - e^{- q l})/\kappa_0 q$ is the two-dimensional Fourier transform of Coulomb potential, $\kappa_0$ is the background dielectric constant, and $l$ is distance to the gate. With the neglect of electron-hole collisions, the system of generalized hydrodynamic equations with self-consistent field is reduced to
\begin{equation}
F_{ac}\left( \begin{matrix}
1+ {\chi_e}V_0(q) & {\chi_e}  \\
{\chi_h} & 1+ {\chi_h}V_0(q)  \\
\end{matrix} \right)\left( \begin{aligned}
&  n_e \\ 
&  n_h \\ 
\end{aligned} \right)=\left( \begin{aligned}
& 0 \\ 
& 0 \\ 
\end{aligned} \right),
\end{equation}
where 
\begin{gather}
F_{ac} = 1- i\eta\mathcal I_2 + \frac{2\eta^2 \mathcal I_1^2}{1-i\eta \mathcal I_0},\\
\chi_e = - 2 n_e \frac{qv_0 \tau_e \eta \mathcal I_1}{1-i\eta \mathcal I_0}\left[\frac{1}{M_{e,k}} + \frac{1-F^{-1}_{ac}}{M_{e,hd}}\right],
\end{gather}

\begin{figure}[t]
	\includegraphics[width=0.9\linewidth]{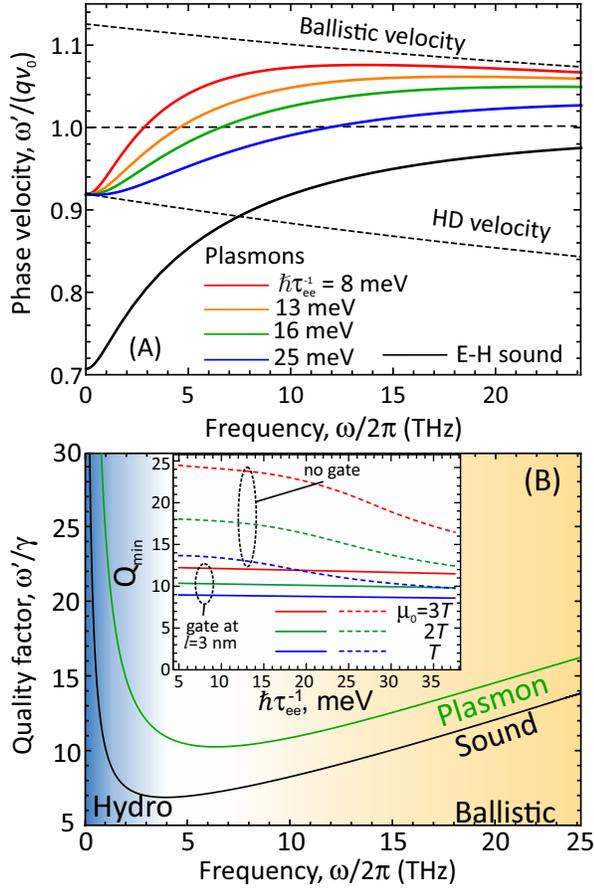}
	\caption{\label{Figure_Velocity_and_QFactor} 
		(A) Phase velocity of collective excitations in graphene with metal gate at  $l=3$ nm, Fermi energy $\mu_0 = 50$~meV, $T=300$ K, dielectric constant $\kappa_0 = 5$. Curves are plotted at various e-e collision frequencies dictating the position of HD-to-ballistic crossover. (B) Quality factor vs frequency. Blue domain corresponds to HD regime with $Q\propto \omega^{-1}$, orange domain corresponds to ballistic regime with constant damping rate $\sim \tau^{-1}_{ee}$ and $Q\propto \omega$. Inset: minimum $Q$-factor in the intermediate regime for gated (solid lines) and non-gated (dashed lines) graphene plasmons
	}
\end{figure}

$i\eta =(1 - i\omega \tau_{ee})^{-1}$. The zeros of function $F_{ac}$ provide the dispersion of quasi-neutral acoustic collective excitations (also referred to as ''electron-hole sound''~\cite{Our-hydrodynamic}, ''energy waves''~\cite{Levitov_EnergyWaves,briskot-nonlinear}, and even ''demons''~\cite{PRL_Demons}). The functions $\chi_e$ and $\chi_h$ are the polarizabilities of electrons and holes, while the equation $1+V_0(q)(\chi_e + \chi_h) = 0$ is the dispersion relation of plasmons. In the HD regime, the dispersion of sound waves is
\begin{equation}
\omega_{\rm sound} \approx \frac{q v_0}{\sqrt 2} - \frac{i}{2}\nu q^2 + \frac{7}{4{\sqrt 2}} \frac{\nu^2 q^4}{qv_0},
\end{equation}
where we have introduced the kinematic viscosity $\nu = v_0^2 \tau_{ee}/4$ (supporting section IV). The damping has a viscous character, and there's no trace of Landau damping despite the sub-quasiparticle mode velocity. This is not unusual as the normal sound in gases also does not exert Landau damping. What is really unexpected that the sound mode remains well-defined even in the formally collisionless regime $\omega\tau_{ee} \gg 1$, where its dispersion reads
\begin{equation}
\label{Sound-ballistic}
\omega_{\rm sound} \approx q v_0 - i \tau^{-1}_{ee} - \frac{9}{2 qv_0\tau_{ee}^2}.
\end{equation}
It is moreover striking if one recalls that intrinsically collisionless models of collective excitations~\cite{Das_Sarma_Plasmons,Ryzhii-plasmons} do not predict any sound modes, though the dispersion (\ref{Sound-ballistic}) falls within their range of validity. The complete picture of sound velocity is shown in Fig. \ref{Figure_Velocity_and_QFactor}, black line: with increasing the wave vector it grows up to the Fermi velocity. One can explain the existence of sound waves in the ballistic domain as a result of strong collisions between carriers with collinear momenta. A similar mode was predicted by summation of vertex corrections in the charge response function~\cite{Gangadharaiah_NovelPlasmonMode}, though the full relation between these results is yet to be established.   

The case of plasmonic excitations is also readily studied within the developed framework. We can parametrize their dispersion law as $\omega = s v_0 q - i \gamma$, where $s$ is the phase velocity in units of Fermi velocity and $\gamma$ is the damping. An expansion of general dispersion relation yields
\begin{equation}
s= \left\{ \begin{aligned}
& \sqrt{\frac{1}{2}+K},\,\,\omega {{\tau }_{ee}}\ll 1 , \\ 
& \frac{1+2K}{\sqrt{1+4K}},\,\,\omega {{\tau }_{ee}}\gg 1, \\ 
\end{aligned} \right.
\end{equation}
where we have introduced the dimensionless ''stiffness'' $K = V_0(q) [n_{0e}/M_e + n_{0h}/M_h]$, kinetic and hydrodynamic masses should be substituted in $K$ in the respective limits. 

The difference between ballistic and hydrodynamic wave velocities is not merely a difference between $M_k$ and $M_{hd}$. While the lower bound for hydrodynamic velocity is the speed of sound $v_0/\sqrt{2}$, the lower bound for ballistic velocity is the speed of quasiparticle $v_0$~\cite{Das_Sarma_Plasmons,Ryzhii-plasmons}, as shown in Fig. \ref{Figure_Velocity_and_QFactor}A. The underlying reason for such behavior is the emergence of a square-root singularity in the conductivity in kinetic regime $\sigma(q,\omega) \propto |q^2v_0^2 - \omega^2|^{1/2}$ which does not let plasmons enter the domain of Landau damping. The difference between two velocities should have profound impact on the possibility to excite plasmons by direct current with Cerenkov mechanism~\cite{Chaplik_AbsorptionEmission2dPlasmons}: it should be possible in the hydrodynamic regime and impossible in ballistic.

Expansion of general dispersion relation also allows to find intrinsic plasmon damping:
\begin{equation}
\gamma =\left\{ \begin{aligned}
& \frac{\nu {{q}^{2}}}{2}+\frac{{{\omega }^{2}}{{\tau }_{ee}}}{8}\left( \frac{{{M}_{hd}}}{{{M}_{k}}}-1 \right){{\left( 2-\frac{1}{{{s}^{2}}} \right)}^{2}},\omega {{\tau }_{ee}}\ll 1, \\ 
& \frac{{{q}^{2}}v_{0}^{2}}{8{{\omega }^{2}}{{\tau }_{ee}}}+\frac{1-{{M}_{k}}/{{M}_{hd}}}{2{{\tau }_{ee}}},\omega {{\tau }_{ee}}\gg 1, \\ 
\end{aligned} \right.
\end{equation} 
which has the viscous ($\propto q^2$) and non-Galilean ($\propto 1-M_k/M_{hd}$) contributions. The effect of collisions on damping in HD and ballistic regimes is the opposite: they give rise to damping in the ballistic regime~\cite{Principi_IntrinsicLifetime} and prevent damping in the HD regime. An unexpected fact is that damping can be small even in the intermediate regime $\omega\tau_{ee} \sim 1$. This is illustrated in Fig.~\ref{Figure_Velocity_and_QFactor}B, where we show the minimum plasmonic quality factor $Q = \omega'/\gamma$ vs. scattering rate for both gated and non-gated plasmons. Both increase in Fermi energy and gate-to-graphene distance lead to reduction in wave vector (at fixed frequency) and decrease in viscous damping, this translates in an elevated $Q$-factor. The non-Galilean damping also disappears with increase in Fermi energy. A large value of minimum intrinsic $Q$-factor $\sim 10-30$ holds the prospective for graphene plasmonics even in the dangerous frequency range $\omega\tau_e \sim 1$.

{\it Electron-hole scattering effects.} 
The developed two-fluid generalized hydrodynamic model displays two independent branches of collective excitations in the absence of e-h scattering: plasma and sound waves. This contrasts to a single-fluid description, where a plasmon gradually transforms into sound wave with reduction of carrier density toward charge neutrality~\cite{Levitov_EnergyWaves,Sun_LinearNonlinear}. 

The account of e-h collisions partly restores the single-fluid picture by leading to strong damping plasmons at charge neutrality point (CNP) and sound waves in degenerate systems. Damping of plasmons occurs due to opposite velocities of oscillating electrons and holes, which causes mutual friction. An estimate for this damping -- which is additive to viscous and ''non-Galilean'' -- can be obtained from (\ref{Gen-continuity}-\ref{Gen-energy}) in the limit $\omega\tau_e \ll 1$, $\{\tau^{tr}_{eh},\tau^{th}_{eh}\}\gg \tau_{e}$:

\begin{figure}[t]
	\includegraphics[width=0.9\linewidth]{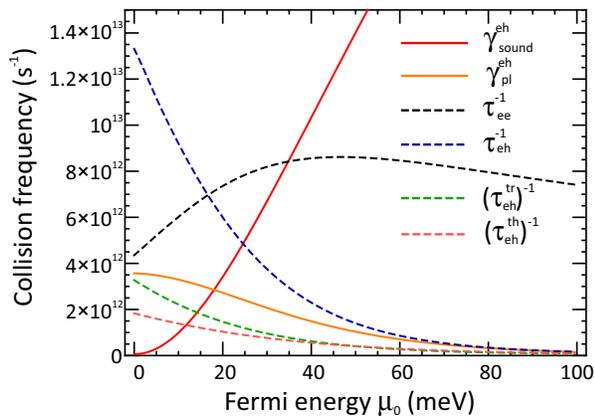}
	\caption{\label{Figure_FrictionalDamping} 
		Damping due to e-h scattering for hydrodynamic collective excitations in graphene vs Fermi energy. Red: electron-hole sound, orange: plasmons. Dashed lines show the characteristic collision frequencies of the model: electron-electron (black), electron-hole (blue), transport (green) and energy (red) scattering rates for e-h collisions. Temperature $T=300$~K, plasmon velocity $s=3.5v_0$, other parameters as in Fig.~\ref{Figure_Velocity_and_QFactor}}
\end{figure}

\begin{gather}
\label{EH-damp-sound}
\gamma^{\rm eh}_{\rm sound} = \frac{(n_{0e}-n_{0h})^2}{n_{0h}^2\varepsilon_{0e} + n_{0e}^2\varepsilon_{0h}}\left[\frac{\varepsilon_{0e}}{2\tau^{tr}_{eh}} + \frac{T_0 \partial\varepsilon_e/\partial T}{2\tau^{th}_{eh}}\right],\\
\label{EH-damp-plasm}
\gamma^{\rm eh}_{\rm pl} = \frac{1}{2}\frac{(M_{e}+M_h)^2}{M^2_h \varepsilon_{0e} + M_e^2 \varepsilon_{0h}}\left[\frac{\varepsilon_{0e}}{\tau^{tr}_{eh}} + \frac{1}{s^2}\frac{T_0 \partial\varepsilon_e/\partial T}{2\tau^{th}_{eh}}\right].
\end{gather}  
These quantities are plotted in Fig. \ref{Figure_FrictionalDamping} as a function of Fermi energy. A rapid increase in sound damping occurs away from CNP because the minority carriers need to oscillate with high velocity to maintain the wave neutrality, which causes strong e-h friction. The damping of plasmons at high Fermi energies drops exponentially. 
Large frictional damping makes the excitation of plasmons at CNP almost impossible, though the commonly accepted mechanism of interband absorption is blocked at $\hbar\omega\ll T$ due to close occupation numbers of initial and final states. At the same time, random fluctuations of carrier density due to residual impurities can cause frictional damping of sound waves in nominally neural system. For typical impurity density $n_i\approx 10^{10}...10^{11}$ cm$^{-2}$, the average fluctuation of Fermi energy is $10...25$ meV~\cite{AdamSelf-Consistent} which corresponds to damping $\gamma^{eh}_{\rm sound} = (1...5)\times 10^{12}$ s$^{-1}$.

In conclusion, we have developed an exactly solvable kinetic model describing both hydrodynamic and non-local ballistic regimes of transport in Dirac materials. The model was used for calculation of conductivity solely due to e-e scattering and studies of collective excitations' properties. We have disentangled the conflicting predictions of hydrodynamic and ballistic models on the spectra and damping of collective excitations, and demonstrated the robustness of novel e-h sound mode at the charge neutrality. The model can be further applied to the studies of thermoelectric properties, which change radically at the diffusive-to-hydrodynamic crossover~\cite{WF-violation-theory,WF-violation-exp}. Future extensions can include different relaxation rates for modes of distribution function; this would capture the low-angle character of Coulomb relaxation in 2d~[\onlinecite{Gurzhi_NewHDEffect}] and can be important for description of transport in confined structures~\cite{Levitov_HeadOn}.

This work was supported by Grant No. 16-19-10557 of the Russian Science Foundation.

\bibliography{Bibliography}

\end{document}